\documentclass{article}

\usepackage{PRIMEarxiv}

\usepackage[utf8]{inputenc} 
\usepackage[T1]{fontenc}    
\usepackage{hyperref}       
\usepackage{url}            
\usepackage{booktabs}       
\usepackage{amsfonts}       
\usepackage{nicefrac}       
\usepackage{microtype}      
\usepackage{lipsum}
\usepackage{fancyhdr}       
\usepackage{graphicx}       
\graphicspath{{media/}}     
\usepackage{amsmath} 

\title{Diffusion of an active particle bound to a generalized elastic model: fractional Langevin equation 
}

\author{Alessandro Taloni\\
 Istituto Sistemi Complessi, Consiglio Nazionale delle Ricerche, via dei Taurini 19, 00185 Rome, Italy\\
  \texttt{alessandro.taloni@isc.cnr.it} \\ 
}

\begin{document}
\maketitle

\begin{abstract}
 We investigate the influence of a self-propelling, out-of-equilibrium active particle on generalized elastic systems, including flexible and semiflexible polymers, fluid membranes, and fluctuating interfaces, while accounting for long-ranged hydrodynamic effects. We derive the fractional Langevin equation governing the dynamics of the active particle, as well as that of any other passive particle (or probe) bound to the elastic system. This equation demonstrates analytically how the active particle dynamics is influenced by the interplay of both  the non-equilibrium force and of the viscoelastic environment. Our study explores the diffusional behavior emerging for both the active particle and a distant probe.The active particle undergoes three different surprising and counterintuitive regimes identified by the distinct  dynamical time-scales: a pseudo-ballistic initial phase, a drastic decrease of the mobility and an asymptotic subdiffusive regime.
\end{abstract}


\section{Introduction}

Active matter refers to a class of materials or systems whose individual components are active, meaning they can convert athermal energy from the environment or internal sources into directed motion or mechanical forces. These systems are characterized by the ability of their constituents to exhibit self-propelled motion, leading to collective behaviors and dynamic patterns that distinguish them from traditional equilibrium systems \cite{bechinger2016active,marchetti2013hydrodynamics, elgeti2015physics, romanczuk2012active, ramaswamy2010mechanics}. 
Thus, by definition,  active systems violate the fluctuation-dissipation theorem (FDT) and encompasses a wide range of phenomena observed in biological systems, such as swarming of birds \cite{reynolds1987flocks,olfati2006flocking,czirok1999collective, bialek2012statistical,cavagna2010scale,attanasi2014information} or schooling of fish \cite{couzin2011uninformed,perez2011collective,couzin2003self}, the run-and-tumble dynamics of microswimmers \cite{berg2004coli,lauga2012dance,matthaus2009coli,wu2000particle,leptos2009dynamics} and the molecular motors-driven transport phenomena inside the cell \cite{gal2010experimental,chen2015memoryless,gal2013particle,weber2015random}. Active matter can also originate synthetically, as seen in systems composed of Janus particles that become active due to chemical reactions \cite{palacci2010sedimentation,howse2007self}, magnetic \cite{dreyfus2005microscopic,tierno2008controlled,ghosh2009controlled} or electrodynamical forces \cite{bricard2013emergence,di2016controlled,yan2016reconfiguring,nishiguchi2018flagellar}. 

\noindent In the past decade, stochastic models have been devised to capture and reproduce observed out-of-equilibrium scenarios and their properties \cite{maggi2017memory,fodor2016far,maggi2014generalized}, as well as the resulting collective dynamics \cite{cates2015motility,bricard2013emergence,reverey2015superdiffusion,budrene1995dynamics,brenner1998physical,cates2015motility,omar2021phase,alert2022active}. These models encompass run-and-tumble particle models \cite{lauga2012dance,kafri2008steady,tailleur2008statistical}, active Brownian particle (ABP) models \cite{ben2015modeling,zheng2013non,howse2007self}, and active Ornstein-Uhlenbeck particle (AOUP) models \cite{nguyen2021active,caprini2022dynamics,sprenger2023dynamics,caprini2022active,wu2000particle,bechinger2016active,samanta2016chain}.

Particular attention has been dedicated to the investigation of active polymer systems, specifically polymeric chains composed of active particles \cite{mackintosh1997actin,eisenstecken2016conformational,kaiser2015does,anand2019behavior} or immersed in an active bath \cite{shin2015facilitation,chaki2019enhanced,kaiser2015does,samanta2016chain,nikola2016active,harder2014activity}.
This interest has been partially motivated by the observed out-of-equilibrium intracellular transport and collective phenomena, where  biopolymers and active elements coexist \cite{weber2012nonthermal,bronshtein2015loss,bronstein2009transient,colin2018actin,sanchez2012spontaneous,kawamura2014chemically,speckner2018anomalous,lin2014structure,mizuno2007nonequilibrium,sonn2017dynamics,sonn2017scale,koster2016actomyosin,celli2005viscoelastic,wagner2017rheological,gan2019mussel,cherstvy2019non,caspi2000enhanced,jeon2011vivo,toyota2011non,henkin2014tunable,kim2019tuning,seisenberger2001real,harada1987sliding,amblard1996subdiffusion,wong2004anomalous,pollard2009actin,chen2015memoryless,song2018neuronal,weihs2006bio,wilhelm2008out,kahana2008active,reverey2015superdiffusion,wang2008rapid,stadler2017non,ku2022effects,yesbolatova2022formulation,vale1988formation}, and partially by the design of new complex active materials.  Usually, simulations of active polymers are enforced by either a self-propelling force tangential to the elastic chain \cite{ghosh2014dynamics,isele2015self,isele2016dynamics,laskar2017filament,chelakkot2012flow}, or by considering the monomers as  active particles, thus resorting to the aforementioned models \cite{eisenstecken2016conformational,kaiser2015does,kaiser2014unusual,liu2019configuration}. The phenomenology exhibited by active polymers is heterogeneous, ranging from translational \cite{jiang2014motion,sarkar2016coarse} to reptation motion \cite{isele2015self,isele2016dynamics}, swelling \cite{eisenstecken2016conformational,kaiser2015does,cao2019crowding}, looping \cite{shin2015facilitation}, swirling \cite{prathyusha2018dynamically} or shrinkage \cite{bianco2018globulelike,duman2018collective}, according to the different parameters characterizing the active polymer model scrutinized.

As emphasized, the numerous numerical studies mentioned above primarily focus on active polymers or polymers embedded in a thermal bath. Nevertheless, three notable studies, both numerical and analytical, specifically investigate the impact of an active particle (or force) confined to a local portion of polymeric chains. In the works by Natali et al. \cite{natali2020local} and Joo et al. \cite{joo2020anomalous}, the dynamics of flexible (Rouse) polymers were examined with the addition of an active monomer, specifically an AOUP. In the first study \cite{natali2020local}, the head of the polymer was permitted to exhibit active non-equilibrium motion. In the second \cite{joo2020anomalous}, the AOUP was positioned as the middle monomer, and the study extended to a flexible polymer network with the central cross-linker being the AOUP. The third study \cite{han2023nonequilibrium}, conceptualized and conducted by part of the team from the second study, concentrated on investigating the impact of an AOUP on a semiflexible polymer network. These works exhibit several common features: $i)$ they are primarily numerical studies, $ii)$ their analysis is rooted in discrete polymer models, and $iii)$ they reveal intriguing and non-trivial physical scenarios, both in the dynamics of the AOUP and in that of the other monomers. Particular interesting are the transition from globule to elongated conformational dynamics observed in \cite{natali2020local}, the intermediate slowing down of the AOUP diffusional motion, in spite of the active self-propelling force, in \cite{joo2020anomalous,han2023nonequilibrium} and the fact that the AOUP dynamics can be described by a fractional Langevin equation \cite{taloni2010generalized}. 

The analysis presented in this paper expands the scope of the applicability of the reported models. Specifically, we explore how an AOUP influences the thermal dynamics of a general viscoelastic system, whether it be a polymer, a membrane, a fluctuating interface, or any other system falling under the category of a generalized elastic model (GEM) \cite{taloni2010generalized, taloni2010correlations}.
The GEM is defined by its stochastic evolution equation: 

\begin{equation}
\frac{\partial}{\partial t}\mathbf{h}\left(\vec{x},t\right)=\int d^dx'\Lambda\left(\vec{x}-\vec{x}'\right) C\frac{\partial^z }{\partial\left|\vec{x}'\right|^z }\mathbf{h}(\vec{x}',t)+\boldsymbol\eta\left(\vec{x},t\right).
\label{GEM}
\end{equation}

\noindent It is formally defined for  $D$-dimensional stochastic field $\mathbf{h}$ defined in the $d$-dimensional infinite space $\vec{x}$.  The GEM encompasses also the presence of hydrodynamic effects, as in the case of a Zimm polymer model by instance \cite{zimm1956dynamics}. 
 The hydrodynamic friction kernel is 

 \begin{equation}
   \Lambda\left(\vec{r}\right)=\frac{B}{\gamma\left|\vec{r}\right|^\alpha}
   \label{hydro}
 \end{equation}

 \noindent where  $\frac{d-1}{2}<\alpha<d$, $B$ is a constant with the dimensions of $L^{\alpha-d}$ and $\gamma$ is the friction constant. The Fourier transform of \eqref{hydro} is $\Lambda\left(\vec{q}\right)
=\frac{(4\pi)^{d/2}}{2^{\alpha}}\frac{\Gamma\left((d-\alpha)/2\right)}{\Gamma\left(\alpha/2\right)}\frac{B}{\gamma}\left|\vec{q}\right|^{\alpha-d}=A\left|\vec{q}\right|^{\alpha-d}
$. In case of local hydrodynamic interactions, \eqref{hydro} simplifies to $\Lambda\left(\vec{r}\right)=\frac{\delta(\vec{r})}{\gamma}$, with $\delta$ the Dirac's delta function.
The fractional derivative appearing in the right hand side of \eqref{GEM} is commonly defined as fractional Laplacian ($\frac{\partial^z}{\partial\left|\vec{x}\right|^z}:=-\left(-\nabla^2\right)^{z/2}$~\cite{kilbas1993fractional}), and is expressed in terms of its Fourier transform ${\cal F}_{\vec{q}}\left\{\frac{\partial^z}{\partial\left|\vec{x}\right|^z}\right\}\equiv-\left|\vec{q}\right|^z$ ~\cite{saichev1997fractional}. Finally the  Gaussian random noise source
satisfies the fluctuation-dissipation (FD) relation

\begin{equation}
    \langle\eta_{j}\left(\vec{x},t\right)\eta_{k}\left(\vec{x}',t'\right) \rangle
=  2k_BT\, l^d\Lambda\left(\vec{x}-\vec{x}'\right)\delta_{j\,k}\delta(t-t'),
\label{FDT}
\end{equation}

\noindent where $j,k\in[1,D]$, $l$ is the microscopical length scale of the model, $k_B$ is the Boltzmann constant and $T$ is the temperature. 

Although the GEM equation \eqref{GEM} is continuous, it simplifies to the Rouse chain model for $z=2$, $d=1$, the semiflexible polymer model for $z=4$, $d=1$, and the Zimm polymer model for $z=2$, $d=1$, and $\alpha=1/2$. The influence of an active non-equilibrium particle on the elastic system described by \eqref{GEM} can be examined using the framework developed in \cite{taloni2011unusual,taloni2013generalized,taloni2016kubo}, where the effect of a local external perturbation was investigated within the context of the Kubo fluctuation relations. It's worth noting that this analysis can be formally extended to the case of out-of-equilibrium stochastic forces.

\subsection{Generalized elastic model with active Brownian particle}

According with the theory outlined in \cite{taloni2011unusual,taloni2013generalized,taloni2016kubo}, it is possible to include in the GEM stochastic evolution \eqref{GEM} the action of an  ABP at a given position $\vec{x}^{\star }$:

\begin{equation}
\frac{\partial}{\partial t}\mathbf{h}\left(\vec{x},t\right)=\int d^dx'\Lambda\left(\vec{x}-\vec{x}'\right) \left[C\frac{\partial^z }{\partial\left|\vec{x}'\right|^z
}\mathbf{h}(\vec{x}',t)+\frac{\gamma\sigma^d}{\gamma_A}\boldsymbol\xi(t)\delta(\vec{x}'-\vec{x}^{\star
})\right]+\boldsymbol\eta\left(\vec{x},t\right).
\label{GEM_potential}
\end{equation}

\noindent Here the active noise $\boldsymbol\xi(t)$ is due to an athermal energy source, leading to the breakdown of the FDT \cite{wu2000particle,maggi2017memory,um2019langevin}. It is governed by the Langevin equation \cite{um2019langevin,eisenstecken2016conformational} 

\begin{equation}
    \frac{\partial\boldsymbol\xi}{\boldsymbol\partial t}=-\frac{\boldsymbol\xi(t)}{\tau_A}+\sqrt{\frac{2\gamma_A\nu_p}{3\tau_A}}\boldsymbol\lambda(t)
    \label{AOUP-LE}
\end{equation}

\noindent where $\sigma$, $\gamma_A$, $\nu_p$ and $\tau_A$ are the characteristic length, the frictional coefficient, the propulsion velocity and the correlation times of the Brownian active particle. The zero-mean Gaussian noise in Eq.\eqref{AOUP-LE} satisfies the fluctuation-dissipation relation $\langle\lambda_\mu(t)\lambda_\nu(t')\rangle=\delta_{\mu,\nu}\delta(t-t')$ ($\mu,\nu\in[1,D]$). The Langevin Eq.\eqref{AOUP-LE} rules the dynamics of the active Ornstein-Ulhenbeck (OU) noise \cite{um2019langevin,maggi2017memory,wu2000particle} with $\langle \xi(t)\rangle=0$ and exponential decaying autocorrelation function 

\begin{equation}
    \langle\xi_\mu(t)\xi_\nu(t')\rangle = \frac{\gamma^2\nu_p^2}{3}\delta_{\mu,\nu}e^{-\frac{|t-t'|}{\tau_A}}.
    \label{AOUP-CF}
\end{equation}

\noindent Hence, hereafter, we will refer to the self-propelling particle at $\vec{x}^{\star
}$ as the AOUP.   The Eq.\eqref{GEM_potential}  jibes with that furnished in \cite{joo2020anomalous}, provided that $C=kl^2$, where $k$ is the flexible polymer's elastic constant. On the other side,  the corresponding  GEM equation for AOUP in a semiflexible chain is obtained by setting  $\Lambda\left(\vec{r}\right)=\frac{\delta(\vec{r})}{\gamma}, z=4, d=1$ \cite{taloni2010correlations}. It is possible to reconcile the evolution equation \eqref{GEM_potential} with that introduced in \cite{han2023nonequilibrium}, assuming  
 that $C=k_BT l_p l$, where $l_p$ is the persistent length of the semiflexible polymer.

 As anticipated, we demonstrate that Eq.\eqref{GEM_potential} constitutes the more general and suitable framework to study the effect of a self-propelling AOUP on a systems whose interactions are non-local and linear, and possibly mediated by long-ranged hydrodynamics. In Sec.\ref{sec:FLE} we derive the fractional Langevin equation for the AOUP and for any particle belonging to the elastic system at a generic position $\vec{x}$, hereafter called probe.
 In Sec.\ref{sec:autoCF} we derive the position  autocorrelation function within the FLE framework. In Sec.\ref{sec:MSD} we furnish the analytical derivation of the mean squared displacement of the AOUP and of the different regimes attained. Moreover we  describe qualitatively the effect of the OU noise on the other regions (or probes) belonging to the elastic system and far from the AOUP. In Sec.\ref{sec:conclusion} we end up with concluding remarks.

\section{Fractional Langevin equation  \label{sec:FLE}}

We firstly furnish the solution of the Eq.\eqref{GEM_potential}  in the Fourier space. We first define the Fourier transform in space and time as  

\begin{equation}
    \mathbf{h}\left(\vec{q},\omega\right)=\int_{-\infty}^{+\infty}d^dx \int_{-\infty}^{+\infty}dt\,
\mathbf{h}\left(\vec{x},t\right)\,e^{-i\left(\vec{q}\cdot\vec{x}-\omega t\right)}\label{FT}
\end{equation}

\noindent  The solution of Eq.(\ref{GEM_potential}) reads

\begin{equation}
\mathbf{h}\left(\vec{q},\omega\right)=\frac{A\sigma^d\gamma}{\gamma_A}\frac{\boldsymbol{\xi}(\omega)e^{-i\vec{q}\cdot\vec{x}^\star}}{\left|\vec{q}\right|^{d-\alpha}\left(-i\omega+AC\left|\vec{q}\right|^{z+\alpha-d}\right)}+\frac{\boldsymbol{\eta}\left(\vec{q},\omega\right)}{-i\omega+AC\left|\vec{q}\right|^{z+\alpha-d}}.
\label{sol_FF}
\end{equation}

\noindent We then multiply both sides of the former equation by $K^+(-i\omega)^\beta$ where 

\begin{equation}
    \beta=\frac{z-d}{z+\alpha-d}
    \label{beta}
\end{equation}

\noindent and

\begin{equation}
K^+= \pi^{d/2-1}\sin\left(\pi\beta\right)\frac{\Gamma(d/2)}{2^{2-d}}\frac{(z+\alpha-d)}{A^\beta C^{\beta-1}},
\label{k+}
\end{equation}

\noindent achieving

\begin{equation}
K^+(-i\omega)^\beta\mathbf{h}\left(\vec{q},\omega\right)=\frac{\sigma^d\gamma}{\gamma_A}\frac{\boldsymbol{\xi}(\omega)\,K^+A(-i\omega)^\beta e^{-i\vec{q}\cdot\vec{x}^\star}}{\left|\vec{q}\right|^{d-\alpha}\left(-i\omega+AC\left|\vec{q}\right|^{z+\alpha-d}\right)}+\frac{\boldsymbol{\eta}\left(\vec{q},\omega\right)K^+(-i\omega)^\beta}{-i\omega+AC\left|\vec{q}\right|^{z+\alpha-d}}.
\label{FLE-FF}
\end{equation}

In analogy to \cite{taloni2011unusual,taloni2013generalized,taloni2016kubo} we define the force-propagator in the Fourier space as

\begin{equation}
    \Theta\left(\left|\vec{q}\right|,\omega\right)=K^+ A\frac{(-i\omega)^\beta e^{-i\vec{q}\cdot\vec{x}^\star}}{\left|\vec{q}\right|^{d-\alpha}\left(-i\omega+AC\left|\vec{q}\right|^{z+\alpha-d}\right)}.
    \label{force_prop-FF}
\end{equation}

\noindent Inverting in space and time, the force-propagator reads \cite{taloni2013generalized,taloni2016kubo}

\begin{equation}
    \Theta\left(\left|\vec{x}-\vec{x}^\star\right|,t\right)=\frac{AK^+\left|\vec{x}\right|^{1-d/2}}{(2\pi)^{d/2}}\int_{0}^{+\infty}d\left|\vec{q}\right|\left|\vec{q}\right|^{\alpha-d/2}J_{d/2-1}\left(\left|\vec{q}\right|\left|\vec{x}-\vec{x}^\star\right|\right)_{-\infty}D_t^{\beta}\left(e^{-AC\left|\vec{q}\right|^{\gamma/2}t}\theta(t)\right)
\label{force_prop},
\end{equation}

\noindent where $\theta(t)$ is the Heaviside step function, $J_{d/2-1}$ is the Bessel function of fractional order $d/2-1$ and the pseudo-differential operator

\begin{equation}
_aD_t^{\beta}\phi(t)=\frac{1}{\Gamma\left(1-\beta\right)}\frac{d}{dt}\int_{a}^tdt'\frac{1}{\left(t-t'\right)^{\beta}}\phi\left(t'\right),
\     \ 0<\beta<1, \label{RL}     
\end{equation}

\noindent represents the left side Riemann-Liouville derivative with lower bound $a<t$ ~\cite{samko1993fractional,podlubny1998fractional}.

\noindent We  also introduce the noise-propagator as 

\begin{equation}
    \label{noise_prop-FF}
    \Phi\left(\left|\vec{q}\right|,\omega\right)=K^+ \frac{(-i\omega)^\beta }{-i\omega+AC\left|\vec{q}\right|^{z+\alpha-d}},
\end{equation}
    
\noindent which, once inverted,  attains the following form in space and time

\begin{equation}
    \Phi\left(\left|\vec{x}\right|,t\right)=\frac{AK^+\left|\vec{x}\right|^{1-d/2}}{(2\pi)^{d/2}}\int_{0}^{+\infty}d\left|\vec{q}\right|\left|\vec{q}\right|^{d/2}J_{d/2-1}\left(\left|\vec{q}\right|\left|\vec{x}\right|\right)_{-\infty}D_t^{\beta}\left(e^{-AC\left|\vec{q}\right|^{\gamma/2}t}\theta(t)\right)
\label{noise-prop}.
\end{equation}

 \noindent Thanks to the definitions \eqref{force_prop-FF} and \eqref{noise_prop-FF}, we can formally invert the Eq.\eqref{FLE-FF}:

 \begin{equation}
     K^+\,_{-\infty}D_t^{\beta}\mathbf{h}\left(\vec{x},t\right)=
\frac{\sigma^d\gamma}{\gamma_A}\int_{-\infty}^{t}dt'\boldsymbol{\xi}(t')\Theta\left(\left|\vec{x}-\vec{x}^\star\right|,t-t'\right)+\boldsymbol{\zeta}\left(\vec{x},t\right) ,
\label{FLE_x} 
 \end{equation}

\noindent where the non-Markovian noise, defined as

\begin{equation}
    \boldsymbol{\zeta}\left(\vec{x},t\right)=\int_{-\infty}^{+\infty}d\vec{x}'\int_{-\infty}^{t}dt'\boldsymbol{\eta}\left(\vec{x}',t'\right)\Phi\left(\left|\vec{x}'-\vec{x}\right|,t-t'\right)
\label{FGN},
\end{equation}

\noindent fulfills the generalized fluctuation-dissipation relation

\begin{equation}
    \langle \zeta_\mu\left(\vec{x},t\right)
\zeta_\nu\left(\vec{x},t'\right)\rangle=k_BT \delta_{\mu,\nu}\frac{K^+}{\Gamma\left(1-\beta\right)\left|t-t'\right|^{\beta}}.
\label{FLE-FDT}
\end{equation}

\noindent The Eq.\eqref{FLE_x} is the fractional equation governing the dynamics of any probe placed at the generic position $\vec{x}$. By a close inspection, its structure unveils how the action of the AOUP in the position $\vec{x}^*$ affects the stochastic behavior of a distant point $\vec{x}$, through the propagator \eqref{force_prop}. This is mirrored by the fGN definition \eqref{FGN}, resulting from the convolution of any stochastic Gaussian force $\boldsymbol{\eta}\left(\vec{x}',t'\right)$, performing  on the entire GEM, with the noise-propagator \eqref{noise-prop}.  Notice that the Riemann-Liouville fractional derivative in \eqref{RL} could be safely replaced  by the Caputo fractional derivative in the FLE \eqref{FLE_x}, as both definitions coincide when the lower bound tends to $-\infty$. \cite{samko1993fractional,podlubny1998fractional}

We now turn to the expression of the effective stochastic equation ruling out the motion of the AOUP. For such a purpose we first perform the Fourier inverse transform in space of the expression \eqref{force_prop-FF}:

\begin{equation}
    \Theta\left(\left|\vec{x}-\vec{x}^*\right|,\omega\right)=
\frac{AK^+(-i\omega)^{\beta}\left|\vec{x}\right|^{1-d/2}}{(2\pi)^{d/2}}\int_{0}^{+\infty}d\left|\vec{q}\right|\frac{\left|\vec{q}\right|^{\alpha-d/2}J_{d/2-1}\left(\left|\vec{q}\right|\left|\vec{x}-\vec{x}^*\right|\right) }{-i\omega+AC\left|\vec{q}\right|^{z+\alpha-d}}  ,
\label{force_prop-F}
\end{equation}

\noindent hence we use the expansion of the Bessel function for small arguments as  ~\cite{abramowitz1988handbook}

\begin{equation}
    J_{d/2-1}(r)\sim\frac{1}{\Gamma(d/2)}\left(\frac{2}{r}\right)^{1-d/2}
\label{Bessel_exp},
\end{equation}

\noindent getting 

\begin{equation}
    \Theta\left(0,\omega\right)\sim
\frac{A^\beta K^+2^{1-d/2}}{(2\pi)^{d/2}\Gamma\left(\frac{d}{2}\right)C^{1-\beta}}\int_{0}^{+\infty}dy\frac{y^{\alpha-1}}{1+y^{z+\alpha-d}}.
\label{force_prop_0-F}
\end{equation}

\noindent Solving the integral \cite{gradshteyn2014table}, we finally obtain that

\begin{equation}
    \Theta\left(0,\omega\right)\sim\frac{1}{2l^d}\delta(t-t')
    \label{noise_prop-AOUP}.
\end{equation}

\noindent Therefore the Eq.\eqref{FLE_x} reduces to the following fractional equation for the AOUP:

 \begin{equation}
     K^+\,_{-\infty}D_t^{\beta}\mathbf{h}\left(\vec{x}^*,t\right)=
\frac{\sigma^d\gamma}{2l^d\gamma_A}\boldsymbol{\xi}(t) + \boldsymbol{\zeta}\left(\vec{x}^*,t\right).
\label{FLE_xstar} 
 \end{equation}

\noindent Thus, it is immediately seen how the active noise applies to the AOUP as a \emph{bare} stochastic force. Similar equations were conjectured in Ref.s\cite{joo2020anomalous} and \cite{han2023nonequilibrium}, while analyzing the case of flexible and semiflexible polymers under the action of an ABP. We hereby offered the rigorous derivation in the more general case represented by the GEM \eqref{GEM}. Moreover we established the formal validity of the FLE framework for any particle, not only for the AOUP.

Most importantly, Eqs.\eqref{FLE_x} and \eqref{FLE_xstar} highlight how the motion of any probe in the elastic system, being the AOUP at $\vec{x}^*$ or another at a position $\vec{x}$, corresponds to a fractional Brownian motion \cite{mandelbrot1968fractional}.

\section{$h$-autocorrelation function\label{sec:autoCF}}

The MSD of any particle belonging to the GEM is affected by the presence of the active noise $\boldsymbol{\xi}(t)$. As a matter of fact the general expression for the MSD reads

\begin{equation}
\langle\delta^2h\left(\vec{x},t\right)\rangle\equiv\langle\left[h\left(\vec{x},t\right)-h\left(\vec{x},0\right)\right]^2\rangle= 2\left[\langle h^2\left(\vec{x},t\right)\rangle-\langle h\left(\vec{x},t\right)h\left(\vec{x},0\right)\rangle\right]
\label{MSD_def}
\end{equation}

\noindent where $h\left(\vec{x},t\right)$ represents one of the components of the stochastic field  $\mathbf{h}\left(\vec{x},t\right)$. From the definition \eqref{MSD_def} it turns out that the calculation of the MSD, as of any other physical observable, needs the explicit expression of the two times  autocorrelation function $\langle h\left(\vec{x},t\right)h\left(\vec{x},t'\right)\rangle$. 

\noindent Starting from the single component solution of the  Eq.\eqref{FLE_x} in the Fourier space 

\begin{equation}
   h\left(\vec{x},\omega\right)=
\frac{\sigma^d\gamma}{\gamma_A}\frac{\xi(\omega)\Theta\left(\left|\vec{x}-\vec{x}^\star\right|,\omega\right)}{K^+(-i\omega)^\beta}+\frac{\zeta\left(\vec{x},\omega\right)}{K^+(-i\omega)^\beta},
\label{FLE_x-F}  
\end{equation}

\noindent we can write down

\begin{multline}
    \langle h\left(\vec{x},\omega\right)h\left(\vec{x},\omega'\right)\rangle = 
    \left(\frac{\sigma^d\gamma}{\gamma_A}\right)^2 \left|\vec{x}-\vec{x}^\star\right|^{2-d}\int_{0}^{+\infty}d\left|\vec{q}\right|\left|\vec{q}\right|^{\alpha-d/2}J_{d/2-1}\left(\left|\vec{q}\right|\left|\vec{x}-\vec{x}^*\right|\right) \times\\
     \int_{0}^{+\infty}d\left|\vec{q}'\right|\left|\vec{q}'\right|^{\alpha-d/2}J_{d/2-1}\left(\left|\vec{q}'\right|\left|\vec{x}-\vec{x}^*\right|\right)\frac{\langle\xi(\omega)\xi(\omega')\rangle}{\left(-i\omega+AC\left|\vec{q}\right|^{z+\alpha-d}\right)\left(-i\omega'+AC\left|\vec{q}'\right|^{z+\alpha-d}\right)}+ \\
     \frac{\langle\zeta(\vec{x},\omega)\zeta(\vec{x},\omega')\rangle}{K^{+2}(-i\omega)^\beta(-i\omega')^\beta},
    \label{auto_CF_x-F}
    \end{multline}

\noindent where we made use of the force-propagator Fourier transform \eqref{force_prop-F}. The first addendum on the RHS of the previous equation can be further simplified by resorting to the active noise correlation properties in Fourier space, i.e.

\[
\langle\xi(\omega)\xi(\omega')=\frac{4\pi\nu_p^2\gamma_A^2}{3\tau_A}\frac{\delta(\omega+\omega')}{\left(\frac{1}{\tau_A}\right)^2+\omega^2}
\].

\noindent The second addendum is the usual term accounting for the correlations inherent to the GEM \cite{taloni2010correlations}.
Therefore  the $h-$autocorrelation function gets the final form

\begin{multline}
    \langle h\left(\vec{x},t\right)h\left(\vec{x},t'\right)\rangle = 
    \frac{(\sigma^d\gamma A\nu_p)^2 \left|\vec{x}-\vec{x}^\star\right|^{2-d}}{2^{d-1}\pi^{d+1}3\tau_A}\times\\
    \int_{0}^{+\infty}d\left|\vec{q}\right|\left|\vec{q}\right|^{\alpha-d/2}J_{d/2-1}\left(\left|\vec{q}\right|\left|\vec{x}-\vec{x}^*\right|\right)
     \int_{0}^{+\infty}d\left|\vec{q}'\right|\left|\vec{q}'\right|^{\alpha-d/2}J_{d/2-1}\left(\left|\vec{q}'\right|\left|\vec{x}-\vec{x}^*\right|\right)\times\\
     \int_{-\infty}^{\infty}d\omega
     \frac{e^{-i\omega(t-t')}}{\left[\left(\frac{1}{\tau_A}\right)^2+\omega^2\right]\left(-i\omega+AC\left|\vec{q}\right|^{z+\alpha-d}\right)\left(-i\omega+AC\left|\vec{q}'\right|^{z+\alpha-d}\right)} +\\
     \frac{2^{2-d}k_BT l^d A^\beta C^{\beta-1}}{(z+\alpha-d)\pi^{d/2}\Gamma\left(\frac{d}{2}\right)\cos\left(\frac{\pi\beta}{2}\right)}\int_0^\infty d\omega\frac{\cos[\omega(t-t')]}{\omega^{1+\beta}},
    \label{auto_CF_x}
\end{multline}

\noindent The two terms appearing on the RHS of the autocorrelation function have a different origin and different behaviors. While in the first one it appears clear the dependence on the internal coordinate $\vec{x}$ as well as the absence of any divergence in the $\omega$ space, the second does not depend on the specific internal position but diverges as $|\omega|^{-(1+\beta)}$ in the limit of $\omega\to 0$ \cite{taloni2010correlations}. Therefore, any physically measurable quantity must be organized in a manner that ensures the cancellation of this divergence.

The $h-$autocorrelation function of the AOUP is obtained from Eq.\eqref{auto_CF_x} using the property \eqref{Bessel_exp}

\begin{multline}
    \langle h\left(\vec{x}^*,t\right)h\left(\vec{x}^*,t'\right)\rangle = \left(\frac{\sigma^d\gamma \nu_p}{l^dK^+}\right)^2\frac{1}{ 6\pi\tau_A}\int_0^\infty d\omega\frac{\cos[\omega(t-t')]}{\omega^{2\beta}\left[\left(\frac{1}{\tau_A}\right)^2+\omega^2\right]}+\\
    \frac{2^{2-d}k_BT l^d A^\beta C^{\beta-1}}{(z+\alpha-d)\pi^{d/2}\Gamma\left(\frac{d}{2}\right)\cos\left(\frac{\pi\beta}{2}\right)}\int_0^\infty d\omega\frac{\cos[\omega(t-t')]}{\omega^{1+\beta}}
\label{auto_CF_xstar}
\end{multline}

\noindent The same expression is achievable from the solution of the Eq.\eqref{FLE_xstar} in Fourier space:

\begin{equation}
   h\left(\vec{x}^*,\omega\right)=
\frac{\sigma^d\gamma}{2l^d\gamma_A}\frac{\xi(\omega)}{K^+(-i\omega)^\beta}+\frac{\zeta\left(\vec{x}^*,\omega\right)}{K^+(-i\omega)^\beta},
\label{FLE_xstar-F}  
\end{equation}

\section{Mean Square Displacement \label{sec:MSD}}

By substituting the expression \eqref{auto_CF_x} into the MSD definition \eqref{MSD_def}, it becomes evident that the MSD is composed of the sum of two contributions:

\begin{equation}
    \langle\delta^2h\left(\vec{x},t\right)\rangle =\langle\delta^2h\left(\vec{x},t\right)\rangle_{OUN} +\langle\delta^2h\left(\vec{x},t\right)\rangle_{fGN}.
    \label{MSD_def_1}
\end{equation}

\noindent  As noticed in Ref.\cite{han2023nonequilibrium}, this superposition makes non-trivial the possible diffusive scenario arising when we are consdering both the AOUP and a generic probe. We consider the case of high P\'eclet number where the active dynamics is definitely larger that the thermal counterpart. Indeed, the first in Eq.\eqref{MSD_def_1} arises from the action of the non-equilibrium Ornstein-Uhlenbeck noise ruled by the Langevin equation \eqref{AOUP-LE}, whereas the second stems from the fractional Gaussian noise \eqref{FGN} and represents the typical subdiffusive thermal dynamics exhibited by any element belonging   to the GEM \eqref{GEM} \cite{taloni2010correlations,taloni2010generalized}:

\begin{equation}
    \langle\delta^2h\left(\vec{x},t\right)\rangle_{fGN}= \frac{4k_BT\pi^{d/2}(AC)^{\beta}\Gamma(1-\beta)}{(2\pi)^d\Gamma(d/2)(z-d)C}t^\beta
    \label{MSD_FGN}
\end{equation}

\noindent The first term on the RHS of \eqref{MSD_def_1} is formally defined as 

\begin{multline}
\langle\delta^2h\left(\vec{x},t\right)\rangle_{OUN}= \frac{(\sigma^d\gamma A\nu_p)^2 \left|\vec{x}-\vec{x}^\star\right|^{2-d}}{2^{d-1}\pi^{d+1}3\tau_A}\times\\
\int_{0}^{+\infty}d\left|\vec{q}\right|\left|\vec{q}\right|^{\alpha-d/2}J_{d/2-1}\left(\left|\vec{q}\right|\left|\vec{x}-\vec{x}^*\right|\right)
     \int_{0}^{+\infty}d\left|\vec{q}'\right|\left|\vec{q}'\right|^{\alpha-d/2}J_{d/2-1}\left(\left|\vec{q}'\right|\left|\vec{x}-\vec{x}^*\right|\right)\times\\
     \int_{-\infty}^{\infty}d\omega
     \frac{1-e^{-i\omega t}}{\left[\left(\frac{1}{\tau_A}\right)^2+\omega^2\right]\left(-i\omega+AC\left|\vec{q}\right|^{z+\alpha-d}\right)\left(-i\omega+AC\left|\vec{q}'\right|^{z+\alpha-d}\right)}.
    \label{MSD_OUN}
\end{multline}

\noindent The equivalent term in case of the AOUP can be derived from \eqref{auto_CF_xstar}

\begin{equation}
    \langle \delta^2h\left(\vec{x}^*,t\right)\rangle_{OUN} = \left(\frac{\sigma^d\gamma \nu_p}{l^dK^+}\right)^2\frac{1}{ 3\pi\tau_A}\int_0^\infty d\omega\frac{1-\cos(\omega t)}{\omega^{2\beta}\left[\left(\frac{1}{\tau_A}\right)^2+\omega^2\right]}
\label{MSD_OUN_xstar}
\end{equation}

\noindent and it is definitely easier to treat. Hence, in the following we develop our analysis starting from the expression \eqref{MSD_OUN_xstar}.

\subsection{AOUP's MSD}

Firstly we notice how in the integral 

\[{\cal I}=\int_0^\infty d\omega\frac{1-\cos(\omega t)}{\omega^{2\beta}\left[\left(\frac{1}{\tau_A}\right)^2+\omega^2\right]}\]

\noindent  there is no issue of divergence in the limit as $\omega\to 0$. In fact, the integrand function behaves as $\sim \omega^{2(1-\beta)}$ for small frequencies. We then introduce a change of variable, $y=\omega \tau_A$, so that

\begin{equation}
    {\cal I}=\tau_A^{1+2\beta}\int_0^\infty dy\frac{1-\cos\left(\frac{yt}{\tau_A}\right)}{y^{2\beta}\left(1+y^2\right)}.
    \label{Int}
\end{equation}

\noindent Its time behavior differs whether $t\ll\tau_A$ or $t\gg\tau_A$ and we will treat them separately.

\subsubsection{$t\ll\tau_A$}

The integral \eqref{Int} must be handled differently in the three cases $0<\beta<1/2$, $\beta=1/2$ and $1/2<\beta<1$.

\begin{itemize}
    \item $0<\beta<1/2$.

    We can split the integral and solve the first one \cite{gradshteyn2014table}:

\begin{equation}
    {\cal I}=\tau_A^{1+2\beta}\left\{\frac{\pi}{2}cosec\left[\frac{(1-2\beta)\pi}{2}\right]
    -\int_0^\infty dy\frac{\cos\left(\frac{yt}{\tau_A}\right)}{y^{2\beta}\left(1+y^2\right)}\right\}.
    \label{Int_t<tau_A_beta<1/2_0}
\end{equation}

\noindent Hence we integrate the second by parts and we expand the resulting trigonometric functions for small arguments

\begin{multline}
     {\cal I}=\tau_A^{1+2\beta}\left\{\frac{\pi}{2}cosec\left[\frac{(1-2\beta)\pi}{2}\right]\right.
    -\\
    \left.\frac{1}{1-2\beta}\left[\left(\frac{t}{\tau_A}\right)^2\left( \int_0^\infty dy \frac{y^{2-2\beta}}{
    1+y^2} - \int_0^\infty dy \frac{y^{3-2\beta}}{
    (1+y^2)^2} \right) + 2\int_0^\infty dy \frac{y^{2-2\beta}}{
    (1+y^2)^2}
    \right]\right\}.
    \label{Int_t<tau_A_beta<1/2_1}
\end{multline}

\noindent By evaluating the remaining integrals we obtain the final result
 
\begin{multline}
     {\cal I}=\tau_A^{1+2\beta}\left\{\frac{\pi}{2}\left(\frac{t}{\tau_A}\right)^2\left[\frac{cosec\left[\frac{(2\beta-3)\pi}{2}\right]}{1-2\beta}+\frac{cosec\left[\frac{(1-2\beta)\pi}{2}\right]}{2}     
     \right] + \right.\\
     \left.\pi\left[ \frac{cosec\left[\frac{(1-2\beta)\pi}{2}\right]}{2} - \frac{\beta\, cosec\left[\pi\beta\right]}{1-2\beta} 
     \right]\right\}.
    \label{Int_t<tau_A_beta<1/2_2}
\end{multline}

    \item $\beta=1/2$.

    The integral \eqref{Int} is in this case

\begin{equation}
    {\cal I}=\tau_A^{2}\int_0^\infty dy\frac{1-\cos\left(\frac{yt}{\tau_A}\right)}{y\left(1+y^2\right)}.
    \label{Int_t<tau_A_beta=1/2_0}
\end{equation}

\noindent Integrating by parts we have

\begin{equation}
    {\cal I}=\tau_A^{2}\left\{-\frac{t}{\tau_A}\int_0^\infty\frac{\ln(y)\sin\left(\frac{yt}{\tau_A}\right)}{1+y^2} + 2\int_0^\infty dy \frac{\ln(y)\left[1-\cos\left(\frac{yt}{\tau_A}\right)\right]}{(1+y^2)^2}
   \right\}.
    \label{Int_t<tau_A_beta=1/2_1}
\end{equation}

\noindent We can neglect the second and split the first into two contributions

\begin{equation}
    {\cal I}=\tau_A^{2}\left\{-\frac{t}{\tau_A}\int_0^1\frac{\ln(y)\sin\left(\frac{yt}{\tau_A}\right)}{1+y^2} + -\frac{t}{\tau_A}\int_1^\infty\frac{\ln(y)\sin\left(\frac{yt}{\tau_A}\right)}{1+y^2} 
   \right\}.
    \label{Int_t<tau_A_beta=1/2_2}
\end{equation}

\noindent Then we can retain only the first one, as the second is nearly zero, and expand the sine for small arguments, getting \cite{gradshteyn2014table}

\begin{equation}
    {\cal I}=\tau_A^{2}\left\{-\left(\frac{t}{\tau_A}\right)^2\int_0^1\frac{y\ln(y)}{1+y^2}
   \right\}=t^2\frac{\pi^2}{48}.
    \label{Int_t<tau_A_beta=1/2_3}
\end{equation}

\item $1/2<\beta<1$.

This case is the easier to be handled. Expanding the cosine for small arguments in \eqref{Int} yields

\begin{equation}
    {\cal I}=\frac{t^2}{\tau_A^{1-2\beta}}\frac{1}{2}\int_0^\infty dy\frac{y^{2(1-\beta)}}{1+y^2}=\frac{t^2}{\tau_A^{1-2\beta}}\frac{\pi}{4}cosec\left[\frac{(3-2\beta)\pi}{2}\right]
    \label{Int_t<tau_A_beta>1/2_0}
\end{equation}

\end{itemize}

\subsubsection{$t\gg\tau_A$}

This time limit presents the same symmetry of the previous one, therefore we study the behaviour of ${\cal I}$ in the three cases $0<\beta<1/2$, $\beta=1/2$ and $1/2<\beta<1$.

\begin{itemize}
    \item $0<\beta<1/2$.

    From \eqref{Int_t<tau_A_beta<1/2_0}, after integrating by parts, it is obtained

    \begin{multline}
        {\cal I}=\tau_A^{1+2\beta}\left\{\frac{\pi}{2}cosec\left[\frac{(1-2\beta)\pi}{2}\right]
    -\frac{t}{\tau_A}\frac{1}{1-2\beta}\int_0^\infty dy\frac{\sin\left(\frac{yt}{\tau_A}\right)}{y^{2\beta-1}\left(1+y^2\right)}
    -\right.\\
    \left.\frac{2}{1-2\beta}\int_0^\infty dy\frac{\cos\left(\frac{yt}{\tau_A}\right)}{y^{2\beta-2}\left(1+y^2\right)}\right\}.
    \label{Int_t>tau_A_beta<1/2_0}
    \end{multline}

\noindent The major contributions to the integrals appearing in \eqref{Int_t>tau_A_beta<1/2_0} come from $y\sim 0$, hence ${\cal I}$ may be properly approximated to 

 \begin{multline}
        {\cal I}=\tau_A^{1+2\beta}\left\{\frac{\pi}{2}cosec\left[\frac{(1-2\beta)\pi}{2}\right]
    -\frac{t}{\tau_A}\frac{1}{1-2\beta}\int_0^\infty dy\,y^{1-2\beta}\sin\left(\frac{yt}{\tau_A}\right)
    -\right.\\
    \left.\frac{2}{1-2\beta}\int_0^\infty dy\,y^{2-2\beta}\cos\left(\frac{yt}{\tau_A}\right)\right\}.
    \label{Int_t>tau_A_beta<1/2_1}
    \end{multline}

 \noindent We can thus use the method of summation of improper integrals \cite{hardy2000divergent,taloni2011unusual} to finalize the calculation 

 \begin{multline}
        {\cal I}=\tau_A^{1+2\beta}\left\{\frac{\pi}{2}cosec\left[\frac{(1-2\beta)\pi}{2}\right]
    -\left(\frac{t}{\tau_A}\right)^{2\beta-1}\frac{\Gamma(2-2\beta)\sin[\pi(1-\beta)]}{1-2\beta}
    -\right.\\
    \left.\left(\frac{t}{\tau_A}\right)^{2\beta-3}\frac{\Gamma(3-2\beta)\cos\left[\frac{\pi(3-2\beta)}{2}\right]}{1-2\beta}\right\}.
    \label{Int_t>tau_A_beta<1/2_2}
    \end{multline}

\item $\beta=1/2$.

We recap from the expression \eqref{Int_t<tau_A_beta=1/2_1}, neglecting the second integral on the RHS and retaining only the contributions coming from $y\sim0$ in the first:

\begin{equation}
    {\cal I}=\tau_A^{2}\left\{-\frac{t}{\tau_A}\int_0^\infty\ln(y)\sin\left(\frac{yt}{\tau_A}\right)\right\}.
    \label{Int_t>tau_A_beta=1/2_1}
\end{equation}

\noindent Then we can split the resulting integral into two terms 

\begin{equation}
    {\cal I}=\tau_A^{2}\left\{-\frac{t}{\tau_A}\int_0^1\ln(y)\sin\left(\frac{yt}{\tau_A}\right) +-\frac{t}{\tau_A}\int_1^\infty\ln(y)\sin\left(\frac{yt}{\tau_A}\right)\right\}.
    \label{Int_t>tau_A_beta=1/2_2}
\end{equation}

\noindent and consider only the first one as the second yields almost no contribution. Finally, solving the integral \cite{gradshteyn2014table}:

\begin{equation}
    {\cal I}=\tau_A^{2}\left\{ \gamma +\ln\left(\frac{t}{\tau_A}\right) - Ci\left(\frac{t}{\tau_A}\right)\right\},
    \label{Int_t>tau_A_beta=1/2_3}
\end{equation}

\noindent where $\gamma$ is the Euler-Mascheroni constant and $Ci$ is the cosine integral \cite{abramowitz1988handbook}.

\item $1/2<\beta<1$.

As in the previous situations the main contributions to the integral in \eqref{Int} will arise from $y\sim 0$, hence

\begin{equation}
    {\cal I}=\tau_A^{1+2\beta}\int_0^\infty dy\frac{1-\cos\left(\frac{yt}{\tau_A}\right)}{y^{2\beta}}.
    \label{Int_t>tau_A_beta>1/2_0}
\end{equation}

\noindent By integration by parts it becomes

\begin{equation}
    {\cal I}=\frac{\tau_A^{1+2\beta}}{2\beta-1}\left(\frac{t}{\tau_A}\right)\int_0^\infty dy\,y^{1-2\beta}\sin\left(\frac{yt}{\tau_A}\right)
    \label{Int_t>tau_A_beta>1/2_1}
\end{equation}

\noindent and, using the methods of improper integrals \cite{hardy2000divergent}, the final result is 

\begin{equation}
    {\cal I}=\tau_A^{1+2\beta}\left(\frac{t}{\tau_A}\right)^{2\beta-1}\frac{\Gamma(2-2\beta)}{2\beta-1}\sin\left[\pi(1-\beta)\right]
    \label{Int_t>tau_A_beta>1/2_2}
\end{equation}
\end{itemize}

Now, collecting the expressions achieved in these  subsections, we can wrap  them up  in a unique compact formula: 

\noindent when $t\ll \tau_A$:

\begin{equation}
    \langle \delta^2h\left(\vec{x}^*,t\right)\rangle_{OUN} \simeq \left(\frac{\sigma^d\gamma \nu_p}{l^dK^+}\right)^2\frac{t^2}{ 3\pi\tau_A}
    \left\{\begin{array}{cc}
    \tau_A^{2\beta-1}\frac{\pi}{2}\left[\frac{cosec\left[\frac{(2\beta-3)\pi}{2}\right]}{1-2\beta}+\frac{cosec\left[\frac{(1-2\beta)\pi}{2}\right]}{2}     
     \right] & 0<\beta<1/2 \\
    \frac{\pi^2}{48} & \beta=1/2\\
    \tau_A^{2\beta-1}\frac{\pi}{4}cosec\left[\frac{(3-2\beta)\pi}{2}\right] & 1/2<\beta<1;
\end{array}   \right. 
\label{MSD_OUN_t<tau_A_xstar}
\end{equation}

\noindent when $t\gg \tau_A$:

\begin{equation}
    \langle \delta^2h\left(\vec{x}^*,t\right)\rangle_{OUN} \simeq \left(\frac{\sigma^d\gamma \nu_p}{l^dK^+}\right)^2\frac{1}{ 3\pi\tau_A}
    \left\{\begin{array}{cc}
    \tau_A^{1+2\beta}\frac{\pi}{2}cosec\left[\frac{(1-2\beta)\pi}{2}\right] & 0<\beta<1/2 \\
    \tau_A^{2}\ln\left(\frac{t}{\tau_A}\right) & \beta=1/2\\
    t^{2\beta-1}\,\tau_A^2\frac{\Gamma(2-2\beta)}{2\beta-1}\sin\left[\pi(1-\beta)\right] & 1/2<\beta<1;
\end{array}   \right. 
\label{MSD_OUN_t>tau_A_xstar}
\end{equation}

\begin{figure}
\includegraphics[width=8 cm]{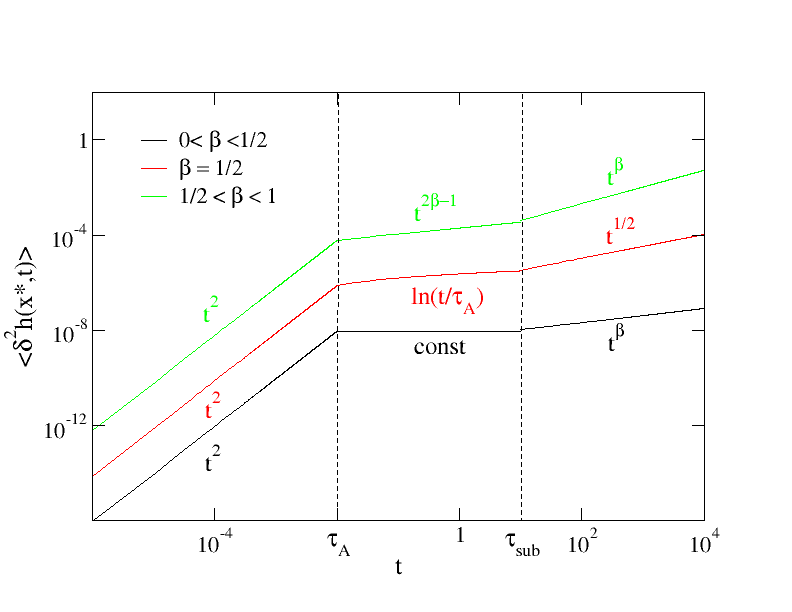}
\caption{MSD of the AOUP. The three situations described in the text, $0<\beta<1/2$ (black curve), $\beta=0$ (red curve) and $1/2<\beta<1$ (green curve) are qualitatively shown. Assuming $\tau_{sub}\gg\tau_A$ the three regimes appear well distinct. After a pseudo- ballistic initial phase, the behaviors in \eqref{MSD_OUN_t>tau_A_xstar} represent a considerable slowing down of the diffusive dynamics, which is followed by the asymptotic subdiffusive GEM usual behaviour \eqref{MSD_FGN}.
\label{fig:1}}
\end{figure}

Hence, we can infer that the impact of the OUN on the AOUP diffusive dynamics exhibits pseudo-ballistic behavior for time intervals shorter than the active decorrelation time ($\tau_A$). However, this impact varies depending on whether $\beta$ is less than, equal to, or greater than 1/2. 
This result must be summed to the contribution arising from the fGn according to the formula \eqref{MSD_def_1}.
It is clear that, in the long time limit, the term $\langle \delta^2h\left(\vec{x}^*,t\right)\rangle_{fGn}$ will dominate. However, the transition from the the dynamics dictated by $\langle \delta^2h\left(\vec{x}^*,t\right)\rangle_{OUN}$ to the asymptotic one will depend on the values of the parameters 
appearing in \eqref{GEM_potential}.  Defining this transition time as $\tau_{sub}$ and assuming $\tau_{sub}\gg\tau_A$ a rough estimate of $\tau_{sub}$ could be given by equating the contributions of the expression \eqref{MSD_OUN_t>tau_A_xstar} with \eqref{MSD_FGN}: $\langle \delta^2h\left(\vec{x}^*,\tau_{sub}\right)\rangle_{OUN}=\langle \delta^2h\left(\vec{x}^*,\tau_{sub}\right)\rangle_{fGn}$. This is schematically shown in Fig.\ref{fig:1}.

\noindent However, it's essential to emphasize that when we relax the assumption of $\tau_{\text{sub}} \gg \tau_A$, the dynamics become less straightforward. Although the asymptotic behavior remains unchanged, the intermediate ultraslow dynamics can be significantly compressed or reduced.

\subsection{MSD at a generic position $\vec{x}$ \label{sec:MSD_x}}

The expression \eqref{MSD_OUN} is not straightforward to manipulate, making it challenging to deduce the impact of the AOUP dynamics on a tracer positioned at a distance of $|\vec{x}-\vec{x}^*|$. The reason is the appearance of a correlation time

\begin{equation}
    \tau(|\vec{x}-\vec{x}^*|)=\frac{|\vec{x}-\vec{x}^*|^{z+\alpha-d}}{CA}
    \label{corr_time},
\end{equation}

\noindent which  can
be seen as the time up to which the dynamics of two distinct probes in $\vec{x}$ and $\vec{x}^*$ is uncorrelated \cite{taloni2011unusual,taloni2012generalized, taloni2013generalized,taloni2016kubo}. As a matter of fact, both the force- and the noise-correlators are expressed as $\Theta\left(\frac{t}{\tau(|\vec{x}|)}\right)$ and $\Phi\left(\frac{t}{\tau(|\vec{x}|)}\right)$. The dependence of \eqref{MSD_def_1} and, in particular, the dependence of the $\langle\delta^2 h(\vec{x},t)\rangle$ on the correlation time \eqref{corr_time} can be achieved by the following changes of variable in the integrals of \eqref{MSD_OUN}: $y=\left(\frac{CA}{-i\omega }\right)^{1/(z+\alpha-d)}|\vec{q}|$, $y'=\left(\frac{CA}{-i\omega }\right)^{1/(z+\alpha-d)}|\vec{q}'|$  and $\lambda=\omega t$: 

\begin{multline}
\langle\delta^2h\left(\vec{x},t\right)\rangle_{OUN}= \frac{(\sigma^d\gamma A\nu_p)^2 }{2^{d-1}\pi^{d+1}3\tau_A}\frac{\left|\vec{x}-\vec{x}^\star\right|^{-\alpha}}{(CA)^{\frac{\alpha}{z+\alpha-d}}}\left(\frac{t}{\tau(\left|\vec{x}-\vec{x}^\star\right|)}\right)^{\frac{2+\alpha-d}{z+\alpha-d}}\int_{-\infty}^{\infty}d\lambda
     \frac{1-e^{-i\lambda }}{\left(\frac{1}{\tau_A}\right)^2+\left(\frac{\lambda}{t}\right)^2}
\times\\
\int_{0}^{+\infty}dy\,\frac{y^{\alpha-d/2}}{1+y^{z+\alpha-d}}J_{d/2-1}\left(
\left(\frac{-i\lambda\tau(\left|\vec{x}-\vec{x}^\star\right|)}{t}\right)^{\frac{1}{z+\alpha-d}}y\right) \times\\
   \int_{0}^{+\infty}dy'\,\frac{y'^{\alpha-d/2}}{1+y'^{z+\alpha-d}}J_{d/2-1}\left(
\left(\frac{-i\lambda\tau(\left|\vec{x}-\vec{x}^\star\right|)}{t}\right)^{\frac{1}{z+\alpha-d}}y'\right).
    \label{MSD_OUN_1}
\end{multline}

As previously mentioned, the formal analysis of \eqref{MSD_OUN} is not straightforward, and it will be the focus of an upcoming investigation. Nonetheless, we can qualitatively examine the limiting behaviors of the equivalent expression \eqref{MSD_OUN_1}. 

\noindent In the regime where $t\ll\tau(|\vec{x}-\vec{x}^*|)$, the Bessel functions exhibit high oscillations. As a result, the major contributions to the integrals come from values where $y\simeq 0$, $y'\simeq 0$, and $\lambda\simeq 0$. Surprisingly, these contributions are almost negligible, leading to:

\begin{equation}
\langle\delta^2h\left(\vec{x},t\right)\rangle\simeq \langle\delta^2h\left(\vec{x},t\right)\rangle_{fGn}.
\label{MSD_x_t<<tau}
\end{equation}

\noindent In the opposite limit, where $t\gg\tau(|\vec{x}-\vec{x}^*|)$, we can employ the expansion of the Bessel functions for small arguments \eqref{Bessel_exp}. This results in the same expression \eqref{MSD_OUN_xstar} as that valid for the AOUP. Consequently, we obtain:

\begin{equation}
\langle\delta^2h\left(\vec{x},t\right)\rangle\simeq \langle\delta^2h\left(\vec{x}^*,t\right)\rangle_{OUN}+\langle\delta^2h\left(\vec{x}^*,t\right)\rangle_{fGn}.
\label{MSD_x_t>>tau}
\end{equation}

Given these limiting situations, we can provide an approximate study of the possible scenarios:

\begin{itemize}
    \item  $\tau(|\vec{x}-\vec{x}^*|)\ll\tau_A$.

     Probes very close to the AOUP exhibit an initial thermal subdiffusive behavior $\propto t^\beta$. Subsequently, the probe at $\vec{x}$ behaves identically to the AOUP.

     \item  $\tau_A\ll\tau(|\vec{x}-\vec{x}^*|)\ll\tau_{sub}$.

     In this case, the probe's diffusion is primarily governed by thermal noise. The non-equilibrium dynamics becomes significant only for $\tau(|\vec{x}-\vec{x}^*|)\ll t\ll \tau_{sub}$, leading to the results in \eqref{MSD_OUN_t>tau_A_xstar}. For longer times, i.e., $t\gg\tau_{sub}$, the thermal MSD \eqref{MSD_FGN} is recovered.

     \item  $\tau_{sub}\ll\tau(|\vec{x}-\vec{x}^*|)$.

     Probes that satisfy this condition, i.e., probes far away from the AOUP, are not influenced by the action of the active force.
    
\end{itemize}

\section{Concluding remarks\label{sec:conclusion}}

In this paper we have examined the diffusional dynamics of an AOUP connected to an elastic system such a (semi)flexible polymer, a membrane, or a fluctuating interface, including hydrodynamic fluid-mediated interactions. Moreover we have investigated the action of the non-equilibrium OU force on the other element belonging to the elastic system. We have demonstrated that the FLE constitutes the correct description of the AOUP dynamics, where the thermal contributions deriving from the rest of the elastic system are incorporated into the fGn, in addition to  a renormalized OUN \eqref{FLE_xstar}. We also have derived the FLE for any other probe placed at an arbitrary distance from the AOUP $|\vec{x}-\vec{x}^*|$. Here the effect of the nonequilibrium force is delayed in time, propagating through the medium thanks to the force-propagator $\Theta(|\vec{x}-\vec{x}^*|,t)$ \eqref{FLE_x}. 

Our analytical theory constitutes a significant improvement over the arguments presented in previous works on this subject. In Ref. \cite{joo2020anomalous}, the FLE for the AOUP was inferred from the numerical evidence of the velocity autocorrelation function's time behavior and from that of the MSD. In other words, it was proposed as an effective equation supported by analytical calculations drawn from a normal mode expansion \cite{ghosh2014dynamics,osmanovic2017dynamics,osmanovic2018properties,sakaue2017active}. However, a formal derivation of the FLE, based on the analysis by Panja \cite{panja2010anomalous,panja2010generalized}, was not attempted.

In Ref. \cite{han2023nonequilibrium}, the mesoscopic FLE for the AOUP attached to a (semi)flexible chain was introduced to reproduce its stochastic non-equilibrium dynamics. The fractional equation, namely the damping kernel, was derived by resorting to the tension propagation theory in the absence of active noise. However, even in this case, a formal derivation from the semiflexible evolution equation was not provided.

In this article, we have offered such an analytical derivation, significantly expanding the domain of its applicability to any elastic system, including hydrodynamics. This became possible because the framework of the FLE with a localized applied potential \cite{taloni2011unusual} could be entirely shifted to the case of the AOUP particle bound to an elastic system. Moreover, this framework entails the derivation of the FLE for any other probe belonging to the GEM, different from the AOUP.

The FLE framework provides a crucial additional value through the formally easy calculation of any observable composed as a function of the elementary correlation function \eqref{auto_CF_x}. In particular, we examined the MSD of AOUP, uncovering some unexpected diffusional scenarios, which include, as special cases, those found in the simulations and theoretical analyses of Refs. \cite{han2023nonequilibrium, joo2020anomalous}.

Firstly, we observed how three different types of diffusion arise depending on the value of $\beta$. This is not immediately apparent in the early time regime, i.e., $t \ll \tau_A$, when the action of the non-equilibrium OU noise drives the directional motion of the AOUP. In this regime, a superdiffusive pseudo-ballistic dynamics emerges, analogous to the result in \cite{joo2020anomalous} where the Rouse model yields $\beta=1/2$ \cite{taloni2010correlations,taloni2010generalized}. However, although the $\sim t^2$ behavior is maintained for any value of $\beta$, the prefactors change significantly as shown in Eq. \eqref{MSD_OUN_t<tau_A_xstar}. Moreover, we observe that the ballistic regime may not distinctly emerge from simulations. Specifically,  at low Péclet numbers, the contribution to the MSD from the thermal part, attributed to the action of the fractional Gaussian noise in \eqref{MSD_def_1}, cannot be entirely neglected. This observation might underlie the $\sim t^{3/2}$ behavior exhibited by the AOUP connected to semiflexible polymers \cite{han2023nonequilibrium}. Simultaneously, this discrepancy could be attributed to the fact that, in the presented simulation results, the AOUP was attached to a network of four semiflexible arms. Regardless, as highlighted by the same authors, in the simulations, 'the superdiffusion for $t\ll\tau_A$ occurs with an anomalous exponent slightly greater than 3/2'.

For times $t\gg\tau_A$ and for high P\'eclet number, the diffusion is still dictated by the non-equilibrium  active force, and the diffusional scenario are different according to the values of $\beta$, see Eq.\eqref{MSD_OUN_t>tau_A_xstar}. For example, we both recover the logarithmic dependence of the Rouse chains, obtained analytically in \cite{joo2020anomalous}, and the $t^{1/2}$ behaviour observed in the numerical simulations  of Ref.\cite{han2023nonequilibrium}. In case of the Zimm polymers model instead, we predict $\sim t^{1/3}$ in this regime, being $\beta=\frac{2}{3}$. Importantly, we anticipate that for $\beta$ in the range $0<\beta<1/2$, the diffusion tends to slow down till to a constant value, see Eq.\eqref{MSD_OUN_t>tau_A_xstar}. 

The asymptotic diffusion requires special attention. The FLEs \eqref{FLE_x} and \eqref{FLE_xstar} involve the superposition of both non-equilibrium and thermal contributions, resulting in the expression \eqref{MSD_def_1}. This implies that, for a high P\'eclet number, there is a crossover between the active subdiffusive motion and the long-time thermal subdiffusive dynamics. The time scale on which this crossover occurs, denoted as $\tau_{sub}$, is not easily determined and crucially depends on the microscopic parameters of the model, as illustrated in Fig.\ref{fig:1}. Situations may arise in which $\tau_{sub}\simeq\tau_A$, causing the contribution in \eqref{MSD_OUN_t>tau_A_xstar} to become less apparent. For the sake of clarity in our analysis, we focus on cases where $\tau_{sub}\gg\tau_A$.

\noindent The transition to thermal motion was also observed in the analysis presented in Ref.s \cite{joo2020anomalous,han2023nonequilibrium}. This crossover time was defined as $\tau_R$, signifying Rouse's time in \cite{joo2020anomalous} and the relaxation time in the case of semiflexible chains \cite{han2023nonequilibrium}. Both interpretations could be seen as thresholds marking the transition to the Browinan linear regime. As we are dealing with infinite systems, $\tau_R\to\infty$ in our context. Consequently, as the influence of the non-equilibrium OU active drive diminishes ($t\gg\tau_A$), the thermal dynamics predominates ($t\gg\tau_{sub}$).

The analysis of the MSD of probes other than AOUP is complex and will be addressed elsewhere. However, even without a detailed analytical derivation, some important conclusions can be drawn. As discussed in Sec.\ref{sec:MSD_x}, the dynamics of regions in the elastic system close to the AOUP are primarily influenced by the active non-equilibrium force up to $\tau_{sub}$, with an initial thermal subdiffusive motion. For $t \gg \tau_{sub}$, the MSD of the probes is still governed by fractional Gaussian noise (fGn).

\noindent The initial thermal regime becomes more pronounced as one considers regions progressively farther from the AOUP. Consequently, the non-equilibrium component of the MSD, denoted as $\langle\delta^2h\left(\vec{x}^,t\right)\rangle_{OUN}$, diminishes. In the limit of very distant regions satisfying $\tau(|\vec{x}-\vec{x}^|)\gg\tau_{sub}$, the diffusion of the probes is entirely dictated by thermal fluctuations, rendering the impact of any non-equilibrium driving force negligible. It's worth highlighting that this diffusional scenario precisely aligns with the findings from the numerical simulations in Ref. \cite{joo2020anomalous}.

 The numerical implementation of GEM with an active particle can be pursued by the numerical integration of the entire system governing Eq.\eqref{GEM_potential}. Alternatively, one could just simulate the AOUP FLE \eqref{FLE_xstar} or the distant probe FLE \eqref{FLE_x}. As stated in \cite{taloni2010generalized} both the description offer the same level of complexity, although the FLE framework provides a simpler and faster way to delve into the dynamical regimes of a single particle.Furthermore, the universal nature of the FLE equation allows for the definition of universality classes, categorized by the value of $\beta$.  The most effective method to simulate the FLE is to utilize the $(k,\omega)$ solution as given in Equation \eqref{FLE-FF} and then invert it.
 
Overall, these results may help to elucidate in vivo dynamics observed in experiments \cite{stadler2017non,speckner2018anomalous}.

\bibliographystyle{unsrt}  
\bibliography{references}

\end{document}